# Spherical coverage verification

Marko D. Petković [1], Dragoljub Pokrajac [2], Longin Jan Latecki [3]

[1] *Faculty of Science and Mathematics, University of Niš,*
*Višegradska 33, 18000 Niš, Serbia*
*E-mail:* `dexterofnis@gmail.com`

[2] *Computer and Information Sciences Department,*
*Applied Mathematics Research Center, CREOSA Center,*
*Delaware State University, Dover, DE, 19901, USA*
*E-mail:* `dpokrajac@desu.edu`

[3] *Computer and Information Sciences Department,*
*Temple University, Philadelphia, PA, USA*
*E-mail:* `latecki@temple.edu`

**Abstract**

We consider the problem of covering hypersphere by a set of spherical hypercaps. This sort of problem has numerous practical applications such as error correcting codes and reverse $k$-nearest neighbor problem. Using the reduction of non degenerated concave quadratic programming (QP) problem, we demonstrate that spherical coverage verification is NP hard. We propose a recursive algorithm based on reducing the problem to several lower dimension subproblems. We test the performance of the proposed algorithm on a number of generated constellations. We demonstrate that the proposed algorithm, in spite of its exponential worst-case complexity, is applicable in practice. In contrast, our results indicate that spherical coverage verification using QP solvers that utilize heuristics, due to numerical instability, may produce false positives.

*Keywords:* geometrical algorithms, quadratic programming, hypersphere, coverage, hypercaps.

## 1 Introduction

We consider the problem of determining whether a hypersphere is completely covered by a set of hyperspherical caps. An equivalent problem is whether the given set of hyperspherical cones centered in the origin, covers the whole space $\mathbb{R}^d$. A three dimensional version of the problem can be solved using the approach from [4], but a solution to a generalized problem in arbitrary dimensional space has not, to the best of our knowledge, been proposed yet. The generalized problem of hyperspherical coverage by a set of hypercaps arises in areas such as coding theory [12] and multidimensional queries [11].

Covering problems are important in computational geometry and have been extensively studied recently. Elbassioni and Tiwary [8] considered the following problem: Given a set of polyhedral hypercones $\mathcal{C}_1, \mathcal{C}_2, \ldots, \mathcal{C}_k$ and a convex set $D \subset \mathbb{R}^d$, check whether cones cover the set $D$ or not. They proved NP completeness in several cases and connected it to the problem of determining whether a union of the convex sets is convex. Also in [3] the authors considered covering a given set of points with a given polygon, whether in [5] authors considered covering the set of points with two disjoint disks and two disjoint squares. Papers [7] and [18] consider covering a sphere (ball) with other spheres (balls).

Recently, development of algorithms for incremental density-based outlier detection [13, 16] have motivated the need to efficiently apply techniques for reverse $k$-nearest neighbor search. The minimal number of hypercaps that can completely cover a hypersphere is related to the theoretical upper bound for the number of reverse $k$-nearest neighbors of a given point and hence the complexity of incremental outlier detection algorithms [15]. Also, the sets of hypercaps that can completely cover the hypersphere are basis for practical algorithms for reverse $k$-nearest neighbors. In this paper, we consider hyperspherical coverage verification: for a given set of hypercaps on a hypersphere we determine whether the set completely covers the hypersphere. We are, however, not concerned how to determine the set of hypercaps that completely covers the hypersphere.

Using the reduction of concave quadratic programming (QP) problem, we demonstrate that spherical coverage verification is NP hard. As a consequence, an algorithm with non-polynomial worst-case complexity may still be viable and practical. We provide a generalized recursive algorithm that can perform coverage verification task for arbitrary dimension $d$. The proposed algorithm is based on reducing the problem to several lower dimension subproblems. In addition, we provide a method that can identify a point on a hypersphere not covered by any hypercap, if such a point exists.

We test the performance of the proposed algorithm on a number of generated constellations with different dimensionality. We demonstrate that the proposed algorithm, in spite of its exponential worst-case complexity, is applicable in practice, with acceptable average-case performance. In contrast, our results indicate that spherical coverage verification using heuristics-based QP solvers, may produce false positives and suffer from numerical instability.

## 2 Spherical coverage verification

In this section, we formally define a spherical coverage verification problem and demonstrate that the considered problem **SphCovVer** can be described as a system of non-linear equations and inequalities. Subsequently, we demonstrate that the problem at hand can be represented as quadratic programming problems with linear constraints.

### 2.1 Problem formulation

Suppose that we have $n$ cones $C_1, \ldots, C_n$ in $d$-dimensional space $\mathbb{R}^d, d \leq 2$ centered at point $O = (0, \ldots, 0)$ and defined by

$$C_i = C(t_i; \theta_i) = \left\{ x \in \mathbb{R}^d \mid \frac{(x, t_i)}{\|x\| \|t_i\|} \geq \theta_i \right\}.$$

Note that each cone $C_i$ is defined by point $t_i \in \mathbb{R}^d$ and real number $-1 < \theta_i < 1$. There holds $\cos \angle xOt_i \geq \theta_i$ for each $x \in C_i$ and $x \neq O$.

For any two given points $x, y \in \mathbb{R}^d$, with $(x, y)$ we denote usual scalar product as $(x, y) = \sum_{i=1}^{d} x_i y_i$ and with $\|x\|$ we denote the Euclidian norm $\|x\| = \sqrt{(x, x)}$.

**Problem 1.** (**SphCovVer**) *Check if cones $C_i$ cover the whole space $\mathbb{R}^d$. Equivalently, check if hypercaps $K_i = C_i \cap S_d(1)$ cover an unit hypersphere $S_d(1) = \{x \in \mathbb{R}^d \mid \|x\| = 1\}$.*

Without loss of generality, we can assume that all points $t_i$ belong to unit hypersphere $S_d(1)$, i.e., which holds $\|t_i\| = 1$. Let $t_i = (t_{i1}, \ldots, t_{id})$. If $x \in S_d(1)$ then:

$$x \in K_i \iff \sum_{j=1}^{d} x_i t_{ij} \geq \theta_i$$

Observe that point $x$ on the unit hypersphere $S_{d-1}(1)$ is not covered by any of the cones $C_1, \ldots, C_n$ if and only if $(x, t_i) < \theta_i$ for all $i = 1, \ldots, n$. Therefore cones $C_1, \ldots, C_n$ cover the unit hypersphere $S_{d-1}(1)$ if and only if the following non-linear system of equations and inequities does not have a solution

$$x_1^2 + x_2^2 + \ldots + x_d^2 = 1$$
$$t_{11}x_1 + t_{12}x_2 + \ldots + t_{1d}x_d < \theta_1$$
$$\vdots$$
$$t_{n1}x_1 + t_{n2}x_2 + \ldots + t_{nd}x_d < \theta_n. \tag{1}$$

## 2.2 Spherical coverage verification as quadratic optimization problem

Denote by $\mathcal{S}$ the solution space of the linear system of inequalities obtained by dropping the first equation in (1). Obviously $\mathcal{S}$ is the convex set. Denote by $\bar{\mathcal{S}}$ the closure of $\mathcal{S}$ i.e., the solution space of the inequalities from (1) when each $<$ is replaced by $\leq$. The set $\bar{\mathcal{S}}$ is also convex. Let $f(x) = \|x\|^2$, $m = f(\tilde{x}) = \min\{f(x) \mid x \in \bar{\mathcal{S}}\}$ and $M = f(x^*) = \max\{f(x) \mid x \in \bar{\mathcal{S}}\}$ (if $\bar{\mathcal{S}}$ is unbounded, then $M = +\infty$). In other words, $\tilde{x}$ and $x^*$ (i.e., $m$ and $M$) are solutions of the following QP problem:

$$(\min/\max) \; f(x) = \|x\|^2 = x_1^2 + x_2^2 + \ldots + x_d^2$$
$$\text{s.t.} \; t_{11}x_1 + t_{12}x_2 + \ldots + t_{1d}x_d \leq \theta_1$$
$$\vdots$$
$$t_{n1}x_1 + t_{n2}x_2 + \ldots + t_{nd}x_d \leq \theta_n. \tag{2}$$

If $\bar{\mathcal{S}}$ is unbounded ($M = +\infty$) then let $x^* \in \bar{S}$ be any feasible point so that $\|x\| > 1$. We say that set $\bar{S}$ specified by constraints from (QP problem (2)) is *degenerated* if it is contained in some hyperplane $\mathcal{H}$. Note that hyperplane $\mathcal{H}$ has to be of the form $(x, t_i) = \theta_i$, i.e. there have to exist two constraints $i$ and $j$ from eq. (2) where $t_{ik} = -t_{jk}$, $k = 1, \ldots, d$ and $\theta_i = -\theta_j$. In such case, the system of equations (1) obviously has no solutions.

The following lemma shows the connection between problem **SphCovVer** (i.e., the system (1)) and the QP problem (2):

**Lemma 1.** *The system of equations and inequities (1) has solutions if and only if $M > 1$, $m < 1$ and $\bar{\mathcal{S}}$ is non-degenerated.*

**Proof.**

($\Leftarrow$:) Let $M > 1$, $m < 1$ and $\bar{\mathcal{S}}$ be non-degenerated. Assume that $x^*$ is a boundary point of $\bar{\mathcal{S}}$ and consider a ball $B_d(x^*, \rho)$ where $\rho < \|x^*\| - 1$. Since $\bar{\mathcal{S}}$ is a non-degenerated polytope, there exists an internal point $x_1^* \in \mathcal{S} \cap B_d(x^*, \rho)$. If $x^*$ is an internal point, we just set $x_1^* = x^*$. Note that $\rho < \|x^*\| - 1$ implies $\|x_1^*\| > 1$.

The same way, we consider a boundary point $\tilde{x}$ of $\bar{\mathcal{S}}$ and construct a new point $\tilde{x}_1 \in \mathcal{S}$ so that' $\|\tilde{x}_1\| < 1$ (here we take $\rho < 1 - \|\tilde{x}\|$).

Since $[\tilde{x}_1, x_1^*] \subset \mathcal{S}$ ($\mathcal{S}$ is convex), function $f(x) = \|x\|^2$ is continuous on $[\tilde{x}_1, x_1^*]$ and $f(\tilde{x}_1) < 1 < f(x_1^*)$, there exists point $u \in (\tilde{x}_1, x_1^*) \subset \mathcal{S}$ so that $f(u) = 1$. In other words, $u$ is the solution of system (1).

($\Rightarrow$:) Let $u$ be one solution of system (1). Hence $u \in \mathcal{S}$ and $f(u) = 1$. Since $\mathcal{S}$ is an open set, there exists a ball $B_d(u; \rho) \subset \mathcal{S}$. Denote by $u_1$ and $u_2$ the intersection points of $B_d(u; \rho)$ and line $Ou$, so that $u \in (Ou_1)$. Obviously holds $f(u_1) = (1 + \rho)^2$ and $f(u_2) = (1 - \rho)^2$. Now $u_1, u_2 \in B_d(u; \rho) \subset \bar{\mathcal{S}}$ and $f(u_1) > 1 > f(u_2)$ directly implies $M \geq f(u_1) > 1 > f(u_2) \geq m$. Non-degeneracy of $\bar{\mathcal{S}}$ follows immediately from the fact that (1) has solutions. □

# 3 Spherical coverage verification is NP-hard

In this section, we prove that the spherical coverage verification problem (**SphCovVer**) is an NP hard problem. First, we demonstrate that the concave non degenerated quadratic programming decision problem (**ConNDQPd**) defined below can be polynomially reduced to **SphCovVer**. Then we demonstrate that the problem **ConNDQPd** is NP complete.

Note that the variant of the problem **ConNDQPd** without the non-degeneracy assumption is considered by Freund and Orlin in [10] (HB problem) where its NP completeness is proven. The degeneracy assumption makes the problem **ConNDQPd** considered here more restrictive than the one considered in [10], implying that we need a different proof of NP completeness. It will be given in the next subsection.

## 3.1 Polynomial reduction of ConNDQPd to SphCovVer

Consider the following concave quadratic programming (QP) problem.

**Problem 2.** (**ConNDQPd**) *Check whether exist* $x_1, x_2, \ldots, x_d \in \mathbb{R}$ *so that*

$$
\begin{aligned}
x_1^2 + x_2^2 + \ldots + x_d^2 &> c \\
a_{11}x_1 + a_{12}x_2 + \ldots + a_{1d}x_d &\leq b_1 \\
&\vdots \\
a_{n1}x_1 + a_{n2}x_2 + \ldots + a_{nd}x_d &\leq b_n.
\end{aligned}
\quad (3)
$$

*where* $c, a_{ij}, b_i \in \mathbb{R}$ *(*$c > 0$*) for* $i = 1, \ldots, n, j = 1, \ldots, d$ *and the polytope specified by* $\leq$ *constraints from* (3) *is* <u>*non-degenerated*</u>.

Also consider the following algorithm:

**Algorithm 1. QP-SphCovVer**
*Input: An instance of the problem* **ConNDQPd**, *i.e., matrix* $A = [a_{ij}] \in \mathbb{R}^{n \times d}$ *and vector* $b = [b_i] \in \mathbb{R}^{n \times 1}$.

1. *Normalize each constraint, i.e compute*

$$
t_{ij} = \frac{a_{ij}}{\sum_{j'=1}^{d} a_{ij'}^2}, \quad \theta_i = \frac{b_i}{c \sum_{j'=1}^{d} a_{ij'}^2}, \quad i = 1, \ldots, n;\ j = 1, \ldots, d. \quad (4)
$$

2. *Solve the minimization problem* (2) *(a convex optimization problem) in polynomial time and denote its minimum by* $m$. *If* $m \geq 1$, *then output* `True`. *If the problem is infeasible, output* `False`. *Otherwise continue.*

3. *Drop each constraint which satisfies* $\theta_i > 1$.

4. *Form the instance of the problem* **SphCovVer** *from the remaining constraints and solve it. Output the complementary result.*

The following theorem proves the correctness of Algorithm **QP-SphCovVer**.

**Theorem 2.** *Algorithm* **QP-SphCovVer** *polynomially reduces the problem* **ConNDQPd** *to the problem* **SphCovVer**.

**Proof.** Note that by (4) we form the equivalent problem of form (2) so that $\|t_i\| = 1$ and $c = 1$.

First assume that $m > 1$. Then all the feasible points satisfy $\|x\|^2 > 1$ and hence the output of problem **ConNDQPd** is `True`. Assume that $m = 1$. By assumption, the feasible set of the problem

(3) is a non-degenerated polytope. Hence it cannot belong to the unit hypersphere (otherwise, it would reduce to the single point) and there is a feasible point $x'$ so that $\|x'\| > 1$. It implies that the answer of **ConNDQPd** is `True`.

Now assume that $m < 1$. Let $\tilde{x}$ be the solution of minimization problem (2). Then the following holds

$$|t_{i1}\tilde{x}_1 + t_{i2}\tilde{x}_2 + \ldots + t_{id}\tilde{x}_d| = |(t_i, \tilde{x})| \leq \|\tilde{x}\|\|t_i\| = \|\tilde{x}\| = m < 1. \tag{5}$$

Equation (5) implies $\theta_i > -1$, since $\tilde{x}$ is a feasible point. Without loss of generality, assume that $\theta_1, \theta_2, \ldots, \theta_p \leq 1$ and $\theta_{p+1}, \theta_{p+2}, \ldots, \theta_n > 1$. We can consider cones

$$C_1 = C(t_1; \theta_1),\ C_2 = C(t_2; \theta_2), \ldots, C_p = C(t_p; \theta_p),$$

since $-1 < \theta_i \leq 1$ and $\|t_i\| = 1$ for all $i = 1, 2, \ldots, p$. Also consider the corresponding system of first $p + 1$ equations from (1):

$$\begin{aligned}
x_1^2 + x_2^2 + \ldots + x_d^2 &= 1 \\
t_{11}x_1 + t_{12}x_2 + \ldots + t_{1d}x_d &< \theta_1 \\
&\vdots \\
t_{p1}x_1 + t_{p2}x_2 + \ldots + t_{pd}x_d &< \theta_p.
\end{aligned} \tag{6}$$

If cones $C_1, C_2, \ldots, C_p$ cover $\mathbb{R}^d$, it implies that the system (6) (and also (1)) does not have a solution. According to Lemma 1, there must hold $M \leq 1$, which implies that the answer of **ConNDQPd** is `False`.

Assume that cones do not cover $\mathbb{R}^d$ and denote by $\bar{x}$ the solution of the system (6). Now, since the following relation holds for $i = p+1, p+2, \ldots, n$:

$$t_{i1}\bar{x}_1 + t_{i2}\bar{x}_2 + \ldots + t_{id}\bar{x}_d = (t_i, \bar{x}) \leq \|\tilde{x}\|\|t_i\| = 1 < \theta_i,$$

we conclude that $\bar{x}$ is also a solution of (1). According to Lemma 1, there holds $M > 1$ and the answer of **ConNDQPd** is `True`. This completes the proof. □

## 3.2 NP completeness of ConNDQPd and SphCovVer

We prove that the problem **ConNDQPd** is NP hard by reducing it to the $k$-clique decision problem. Recall that, for a given graph $G = (V, E)$, set $Cl \subseteq V$ is a *clique*, if for every $u, v \in Cl$ holds $\{u, v\} \in E$. In other words, a clique is every set $Cl$ of vertices, so that each two vertices from $Cl$ are adjacent. Clique $Cl$ is called *k-clique*, if it contains exactly $k$ vertices. The $k$-clique decision problem (see e.g., [6]) can be formulated as follows:

**Problem 3.** ($k$-**Clique**) *Given a graph $G = (V, E)$, check if there exists $k$-clique.*

It is known, ([6]) that the problem $k$-**Clique** is NP complete. The following lemma demonstrates that it can be polynomially reduced to the problem **ConNDQPd**.

**Lemma 3.** *Problem $k$-**Clique** can be polynomially reduced to problem **ConNDQPd**.*

**Proof.** For a given graph $G = (V, E)$ with the vertex set $V = \{1, 2, \ldots, n\}$, consider the following instance of problem **ConNDQPd**:

$$\begin{aligned}
x_1^2 + x_2^2 + \ldots + x_n^2 &> n - \epsilon \\
-1 \leq x_i &\leq 1 \\
x_i + x_j &\leq 0, \quad \forall \{i, j\} \notin E \\
x_1 + x_2 + \ldots + x_n &\geq 2k - n
\end{aligned} \tag{7}$$

Here $0 < \epsilon < 1$ and its value will be determined later. If there exists a clique $Cl$ of length $k$ in graph $G$, then by setting

$$x_i^* = \begin{cases} 1, & i \in Cl \\ -1, & i \notin Cl \end{cases}$$

we obtain one feasible solution $(x_1^*, x_2^*, \ldots, x_n^*)$ of the problem (7) satisfying $(x_1^*)^2 + (x_2^*)^2 + \ldots + (x_n^*)^2 = n > n - \epsilon$.

Now assume that there is no clique of length $k$ in graph $G$. We show that the decision problem (7) does not have the solution. Consider the following auxiliary optimization problem

$$\begin{aligned} \max\ & x_1 + x_2 + \ldots + x_n \\ \text{s.t.}\ & -1 \leq x_i \leq 1, \quad i = 1, 2, \ldots, n \\ & x_i + x_j \leq 0, \quad \forall \{i,j\} \notin E \\ & x_1^2 + x_2^2 + \ldots + x_n^2 > n - \epsilon. \end{aligned} \quad (8)$$

Let $(x_1, x_2, \ldots, x_n)$ be an arbitrary feasible solution. It can be easily checked that $x_i^2 > 1 - \epsilon$ (due to the quadratic condition in (8)) for $i = 1, 2, \ldots, n$ implying that $x_i \in [-1, -\sqrt{1-\epsilon}) \cup (\sqrt{1-\epsilon}, 1]$. Let $x_{i_1}, x_{i_2}, \ldots, x_{i_p} \in (\sqrt{1-\epsilon}, 1]$ and $x_q \in [-1, -\sqrt{1-\epsilon})$ for $q \notin \{i_1, i_2, \ldots, i_l\}$. For arbitrary $1 \leq r < s \leq p$ there holds $\{i_r, i_s\} \in E$, according to $x_{i_r} + x_{i_s} \geq 2\sqrt{1-\epsilon} > 0$. In other words, vertices $i_1, i_2, \ldots, i_p$ form a clique of length $p$ in graph $G$.

According to our assumption that there is no clique of length $k$ in $G$, it must hold that $p < k$. Furthermore it holds that

$$f = x_1 + x_2 + \ldots + x_n = \sum_{j=1}^{p} x_{i_j} + \sum_{q \notin \{i_1, i_2, \ldots, i_p\}} x_q \leq p - (n-p)\sqrt{1-\epsilon} \quad (9)$$

$$< p - (n-p)(1-\epsilon) = 2p - n + \epsilon(n-p).$$

Now by choosing $\epsilon = 2/n$ and by using $k > p$ and (9), we obtain

$$f < 2p - n + \frac{2}{n}(n-p) < 2p - n + \frac{2(k-p)}{n-p}(n-p) = 2k - n.$$

According to the previous expression, each feasible solution of (8) satisfies $f < 2k - n$ implying that the system (7) has no solutions. $\square$

As a direct consequence of the Lemma 3 and the NP completeness of the problem $k$-**Clique**, and since a verification of a solution for the eq. (3) is possible in polynomial time, the following corollary holds:

**Corollary 4.** *Problem* **ConNDQPd** *is NP complete.*

Now Corollary 4 and Theorem 2 directly imply:

**Theorem 5.** *Spherical coverage verification, i.e., the problem* **SphCovVer**, *is NP hard.*

## 4 Algorithms for spherical coverage verification

The simplest method for spherical coverage verification is to apply a non-deterministic Monte-Carlo approach. The idea is to generate a large number $N$ (for example $N = 10^{10}$) of pseudo-random points $x$ distributed uniformly on the sphere. For each generated point $x$ we check if there is hypercap $K_i$ containing the point. If the answer for any point is negative, the algorithm outputs `False`. If the system of hypercaps $K_i$, $i = 1, \ldots, n$ covers the unit sphere, this method always outputs the correct answer `True`.

If there is no coverage, this method will output the correct answer `False` with probability that increases with $N$. However, there is no guarantee that the algorithm will not return false positives (thus providing answer `True` for a non-covering system of hypercaps). Hence, in the subsequent subsections, we discuss algorithms for spherical coverage verification based on application of quadratic programming, and on the reduction of the given problem to lower dimensional subproblems.

## 4.1 QP-based verification

According to Section 2.1, cones $C_i, i = 1, \ldots, n$ cover the space $\mathbb{R}^d$ if an only if the system (1) has no solutions. According to Lemma 1, to check whether the system (1) has solutions, we need to solve the QP problems (2) and to determine whether the minimal value $m$ and maximal value $M$ of the objective function satisfy $m < 1 < M$. The following algorithm for solving problem **SphCovVer** arises from the aforementioned discussion:

**Algorithm 2. Cover-QP**
*Input: Caps $K_i$, $i = 1, \ldots, n$ defined by $t_i \in S_d(1)$ and $\theta_i \in (-1, 1)$.*

1. *Return `True` if QP problems (2) are degenerated. Otherwise, solve both problems.*

2. *If any of the problems is not feasible, or $m < 1 < M$ does not hold, return `True`. Otherwise return `False`.*

To apply algorithm **Cover-QP** we need appropriate QP problem solvers. Note that the minimization problem can be solved in polynomial time, since it is convex. Since concave QP is NP complete (see Section 3.2), one of known heuristics can be applied ([17]).

## 4.2 Recursive algorithm

In this section, we describe our recursive algorithm for solving the problem **SphCovVer**. The main idea is to reduce the initial $d$-dimensional problem to several $d-1$-dimensional problems. More precisely, using d-dimensional inversion, hypercaps are mapped into regions consisting of hyperspheres in $d-1$ dimensional plane and their exteriors/interiors depending on the position of the center of inversion w.r.t. a hypercap. As proven in the next subsection, a $d$-dimensional sphere is covered by the hypercaps if and only if the resulting $d-1$ plane is completely covered by the $d-1$ dimensional regions. In turn, this may be true if a boundary of each region (which is itself a $d-1$ dimensional hypersphere) is completely covered by $d-1$ dimensional hypercaps defined by corresponding regions. Thus, the coverage of $d$-dimensional sphere reduces to coverage of $d-1$ dimensional spheres. The following subsections provide the rationale for the algorithm and discuss the algorithm formally.

### 4.2.1 Rationale of the algorithm

We restate the well-known definition of the inversion in $\mathbb{R}^d$.

**Definition 1.** *Let $c \in \mathbb{R}^d$ be a given point, and let $R$ be a positive real number. Inversion $\psi_{c,R}(x)$ is a function $\psi_{c,R} : \mathbb{R}^d \to \mathbb{R}^d$ that maps every point $x \in \mathbb{R}^d$ to a point $y$ so that:*

$$y = c + \frac{R^2}{\|x - c\|^2}(x - c). \tag{10}$$

*Point $c$ is called the center of inversion $\psi_{c,R}$ and $R$ is the radius of inversion.*

Let us apply inversion $\psi = \psi_{(1,0,\ldots,0),1}$ on caps $K_i$ and unit hypersphere $S_d(1)$. It is well-known that an image of a hypersphere, by inversion whose center belongs to the hypersphere is a hyperplane. Thus, the image of the unit hypersphere is hyperplane $x_1 = 1/2$.

Denote by $\partial X$ the boundary of a given set $X$, Particularly, we denote by $D_i = \partial K_i$ the boundary of cap $K_i$. Also, for a given hypersphere $S$ denote by $\text{int}S$ its interior and by $\text{ext}S$ its exterior. Images of caps $K_i$, $i = 1, \ldots, n$ consist of $d-1$-dimensional hyperspheres, belonging to hyperplane $x_1 = 1/2$ and their exteriors or interiors, depending whether the center of inversion is outside or inside the cap. More precisely, the following Lemma holds:

**Lemma 6.** *Let $c = (1, 0, \ldots, 0) \in \mathbb{R}^d$ and assume that $c \notin D_i$ for every $i = 1, \ldots, n$. Image of $D_i$, by an inversion $\psi_{c,1}$ is a $d-1$-dimensional hypersphere $S_i$ with center $\beta_i = (\beta_{i1}, \ldots, \beta_{id})$ and radius $r_i$. Moreover, the image of $K_i$ is $R_i = S_i \cup \text{ext}S_i$, if $c \in K_i$ and $R_i = S_i \cup \text{int}S_i$, if $c \notin K_i$. Values $\beta_i$ and $r_i$ are given by the following expressions:*

$$\beta_{i1} = \frac{1}{2}, \quad \beta_{ij} = \frac{t_{ij}}{2(\theta_i - t_{i1})}, \quad r_i = \frac{\sqrt{1-\theta_i^2}}{2(\theta_i - t_{i1})}. \tag{11}$$

**Proof.** Translate the coordinate system to the center of inversion, i.e., to point $c = (1, 0, \ldots, 0)$. The equation describing cap $K_i$ becomes ($x' = x - c$):

$$t_{i1}x_1' + \ldots + t_{id}x_d' \geq \theta_i - t_{i1}. \tag{12}$$

Let $\psi(x) = y$. By the involution property of inversion, we can conclude that $\psi(y) = x$ or in other words:

$$x_i' = \frac{1}{\sum_{j=1}^d y_j'^2} y_i'.$$

By replacing the last expression into (12) we obtain:

$$\sum_{j=1}^d t_{ij} y_j' \geq (\theta_i - t_{i1}) \sum_{j=1}^d y_j'^2. \tag{13}$$

Replacing $y = y' + c$ finally yields:

$$R_i : \quad \sum_{j=2}^d (y_j - \beta_{ij})^2 \begin{cases} \geq r_i^2, & \theta_i - t_{i1} > 0 \\ \leq r_i^2, & \theta_i - t_{i1} < 0 \end{cases}, \quad 2 \leq i \leq d, \quad y_1 = \frac{1}{2}. \tag{14}$$

Here condition $\theta_i - t_{i1} > 0$ is equivalent to $c \in K_i$. In (14) we denote:

$$\beta_{i1} = \frac{1}{2}, \quad \beta_{ij} = \frac{t_{ij}}{2(\theta_i - t_{i1})}, \quad r_i^2 = \left(\frac{1}{2(\theta_i - t_{i1})}\right)^2 \sum_{j=2}^d t_{ij}^2 + \frac{t_{i1}(y_1 - 1)}{\theta_i - t_{i1}} - (y_1 - 1)^2. \tag{15}$$

The last expression (15) can be further simplified using $\sum_{j=2}^d t_{ij}^2 = 1 - t_{i1}^2$ and $y_1 = \frac{1}{2}$ into (11). □

In Fig. 1 we illustrate Lemma 6 for three spherical caps $D_i$, $i = 1, 2, 3$ and their images. Observe that $c \in D_3$, hence $D_3$ maps to the exterior of sphere $S_3$.

Due to Lemma 6, problem **SphCovVer** reduces to the following hyperplane cover verification (**HpCovVer**) problem:

**Problem 4.** (**HpCovVer**) *For given $d-1$-dimensional hyperspheres $S_i$, with center $\beta_i$ and radius $r_i$ belonging to the hyperplane $x_1 = 1/2$, and sets $R_i$ so that $R_i = S_i \cup \text{int}S_i$ or $R_i = S_i \cup \text{ext}S_i$ $(i = 1, \ldots, n)$, check if sets $R_i$ cover the whole hyperplane.*

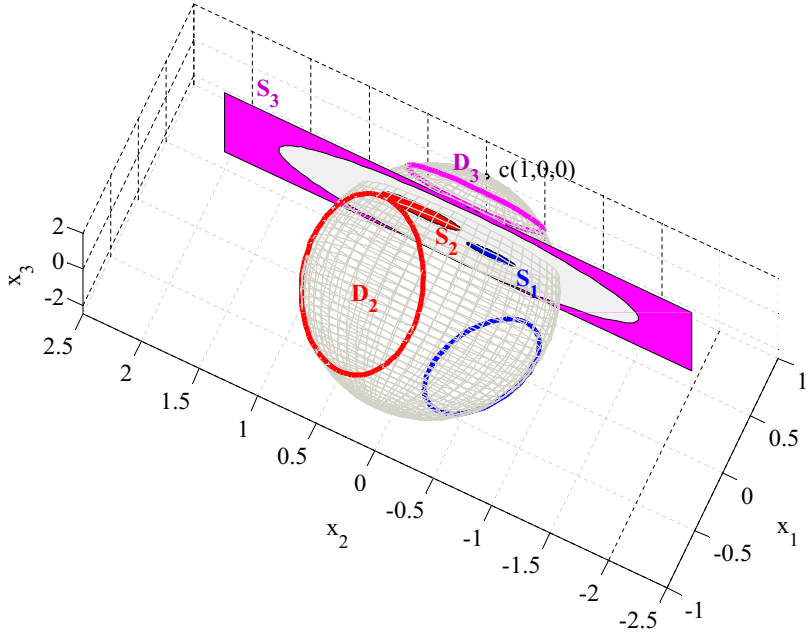

Figure 1: Three spherical caps and their images under the inversion $\psi$ for $d = 3$.

To simplify further discussion, if $R_i = S_i \cup \text{int} S_i$ set $R_i$ is called *internal*; otherwise we call it *external*. If all the sets $R_i$ are internal, then we can immediately conclude that the answer to the problem **HpCovVer** is False. This holds obviously from the fact that all sets $R_i$ are bounded and hence is also their union $\bigcup_{i=1}^{n} R_i$.

Suppose, in contrast, that at least one set $R_i$ is external. If all sets $R$ do not cover the whole hyperplane, there exists one hypersphere $S_i$ and point $x \in S_i$ so that it is uncovered by other sets $R_j$, $j \neq i$. In other words, the following theorem holds:

**Theorem 7.** *Sets $R_1, \ldots, R_n$ with different boundaries $S_i = \partial R_i$ cover the whole hyperplane $x_1 = 1/2$ if and only if every hypersphere $S_i$ is covered by other sets $R_j$, $j \neq i$, i.e.,*

$$S_i = \partial R_i \subset \bigcup_{j \neq i} R_j. \tag{16}$$

**Proof.**

($\Leftarrow$:) If sets $R_1, \ldots, R_n$ cover the whole hyperplane $x_1 = 1/2$, each hypersphere, $S_i$ as a subset of the hyperplane will be covered.

($\Rightarrow$:) Denote an uncovered region of the hyperplane $x_1 = 1/2$ with $Q$. The boundary $\partial Q$ consists of the union of spherical caps. Denote by $A$ one of those caps and by $S_i$ the hypersphere which $A$ belongs to. Since $S_i \neq S_j$ for $i \neq j$, interior $\text{int} A$ cannot be covered by remaining hyperspheres $S_j$, $j \neq i$. In other words, there exists a point $x \in \text{int} A \setminus \bigcup_{i \neq j} S_j$. Point $x$ is covered by some set $R_k$ and since $x \notin S_k$, there holds $x \in \text{int} R_k$. Hence, there exists a ball $B_{d-1}(x; \delta) \subseteq \text{int} R_k$ and holds $B_{d-1}(x; \delta) \cap Q = \emptyset$. This is the contradiction with the fact that $x \in A$ is a boundary point of $Q$.

□

If $S_i = S_j$ for $i \neq j$ then either $R_i$ and $R_j$ cover the hyperplane (if one of them is internal and the other is external) or $R_i = R_j$. In the second case, we can eliminate one of them and continue.

According to Theorem 7 we need to check whether each hypersphere $S_i$ is covered by sets $R_j$, $j \neq i$. We distinguish the following cases, depending on whether the pairs of hyperspheres $S_i$ and $S_j = \partial R_j$ are disjoint:

**Case 1.** Hyperspheres $S_i$ and $S_j$ have a nonempty intersection. In such a case $S_i \cap R_j$ is a hypercap $K_j^i$ defined as:

$$K_j^i = \left\{ x \in S_i \mid \frac{(x - \beta_i, x_j^i)}{\|x - \beta_i\| \|x_j^i\|} \geq \theta_j^i \right\}, \quad i \neq j, \tag{17}$$

where we define:

$$\theta_j^i \triangleq \begin{cases} \frac{r_i^2 + d_{ij}^2 - r_j^2}{2 r_i d_{ij}}, & R_j \text{ is internal} \\ -\frac{r_i^2 + d_{ij}^2 - r_j^2}{2 r_i d_{ij}}, & R_j \text{ is external} \end{cases}, \quad x_j^i \triangleq \begin{cases} d_{ij}^{-1}(\beta_j - \beta_i), & R_j \text{ is internal} \\ -d_{ij}^{-1}(\beta_j - \beta_i), & R_j \text{ is external} \end{cases}, \quad d_{ij} \triangleq \|\beta_i - \beta_j\|, \quad i \neq j. \tag{18}$$

Observe that for the points $x_j^i$ from eq. (18), $x_j^i = ((x_j^i)_2, \ldots, (x_j^i)_d) \in \mathbb{R}^{d-1}$, since $(x_j^i)_1 = 0$ for every $i$ and $j$. Case 1 is illustrated in the Fig. 2.

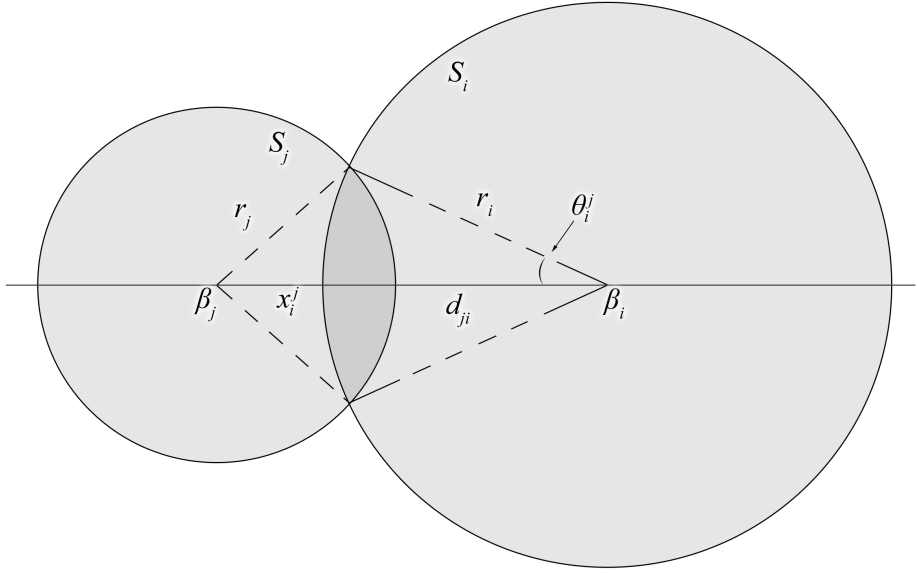

Figure 2: Intersection of $d - 1$-dimensional hyperspheres $S_i$ and $S_j$.

**Case 2.** Hyperspheres $S_i$ and $S_j$ are disjoint. The equivalent condition is $\theta_j^i \notin (-1, 1)$, where $\theta_j^i$ is defined by eq. (18). In such case either $S_i \subset R_j$ (**Case 2a**) or $S_i \cap R_j = \emptyset$ (**Case 2b**) holds. Which of these two sub cases holds can be determined e.g., by choosing the arbitrary point (for example $x_j + (r_j, 0, \ldots, 0)$) on $\partial S_j$ and checking the inequality (14) for hypersphere $S_i$.

For fixed $i$, if for any $j$ the condition $S_i \subset R_j$ (**Case 2a**) is satisfied, then eq. (16) holds and $S_i$ is covered. Therefore, the algorithm may continue with another value of $i$. Otherwise, it is sufficient to determine whether the sphere $S_i$ is covered by those hypercaps $K_j$ defined by eq. (17) and corresponding to the pairs of spheres $(S_i, S_j)$ satisfying **Case 1** (note that pairs $(S_i, S_j)$ satisfying **Case 2b** do not need be considered due to disjointness of $S_i$ and $R_j$). This is an instance of problem **SphCovVer**, for dimension $d - 1$. Hence, we reduce original problem **SphCovVer** to at most $n - 1$ equivalent problems of dimension $d - 1$.

### 4.2.2 Base case of the algorithm

When $d = 2$, the inversion from Lemma 6 maps a $2D$ sphere (a circle) into a straight line and $2D$ caps (arcs) degenerate into intervals. Hence, as the base case, we choose case $d = 2$ of problem **HpCovVer**. We omit the first coordinate (which is equal to $1/2$) of each point from sets $S_i$ and $R_i$. Hence, we assume that there are given sets $R_i$ so that $R_i = [c_i, d_i]$ ($R_i$ is internal) or $R_i = \mathbb{R} \setminus (c_i, d_i)$ ($R_i$ is external) for some real numbers $c_i < d_i$. Here $c_i = \beta_{i2} - r_i$ and $d_i = \beta_{i2} + r_i$. The problem is to check if sets $R_i$ cover the whole real line $\mathbb{R}$, i.e., $\bigcup_{i=1}^{n} R_i = \mathbb{R}$. Without loss of generality we can assume that $R_1, \ldots, R_s$ be external and $R_{s+1}, \ldots, R_n$ internal. Let us define:

$$c' = \max_{i=1,\ldots,s} c_i, \quad d' = \min_{i=1,\ldots,s} d_i.$$

Obviously, if $c' \geq d'$, $\mathbb{R}$ is covered by the sets $R_i$. Otherwise,

$$(c', d') = \mathbb{R} \setminus \bigcup_{i=1}^{s} R_i,$$

and we need to check if the interval $(c', d')$ is covered by $[c_{s+1}, d_{s+1}], \ldots, [c_n, d_n]$. This can be performed by sorting segments $[c_{s+1}, d_{s+1}], \ldots, [c_n, d_n]$ with respect to $c_i$ and sequentially shortening the target interval $(c', d')$. Let us assume that the segments are sorted so that $c_{s+1} \leq c_{s+2} \leq \ldots \leq c_n$. If $c' < c_{s+1}$ then the interval $(c', c_{s+1})$ is uncovered (since $c_{s+1}$ is minimal); hence $\mathbb{R}$ is also uncovered. Otherwise, we need to check if $[c'', d']$, where $c'' = \max\{c', d_{s+1}\}$ is covered by segments $[c_{s+2}, d_{s+2}], \ldots, [c_n, d_n]$. This leads to the following algorithm for solving the base case.

**Algorithm 3. Cover2-2D** ($d = 2$ case of problem **HpCovVer**)
Input: Values $\beta_{i2}$, $r_i$ ($r_i \geq 0$) and $\text{out}_i$ for $i = 1, \ldots, n$. We assume that $\text{out}_i = \texttt{True}$ if $R_i$ is external and otherwise $\text{out}_i = \texttt{False}$.

1. Let $c_i = \beta_{i2} - r_i$ and $d_i = \beta_{i2} + r_i$
2. Reorder the sets $R_i$ so that $\text{out}_i = \texttt{True}$ for $i = 1, \ldots, s$ and $\text{out}_i = \texttt{False}$ for $i = s+1, \ldots, n$.
3. Sort sets $R_i, i = 1 + s, \ldots, n$ so that $c_{s+1} \leq \ldots \leq c_n$.
4. Let $c' := \max_{i=1,\ldots,s} c_i$ and $d' := \min_{i=1,\ldots,s} d_i$
5. For every $i = s+1, \ldots, n$ do the following:
    5.1. If $c' \geq d'$ return `True`. Otherwise continue.
    5.2. If $c_i > c'$ then return `False`. Otherwise set $c' := \max\{c', d_i\}$ and continue.
6. If $c' \geq d'$ return `True`, otherwise return `False`.

Note that the complexity of the Algorithm **Cover2-2D** is $\mathcal{O}(n \log n)$ if an asymptotically optimal sorting algorithm is used for intervals sorting.

### 4.2.3 Algorithm outline

Next, we formulate complete recursive Algorithm **Cover** for solving the general case of problem **SphCovVer**.

**Algorithm 4. Cover**
Input: Caps $K_i$, $i = 1, \ldots, n$ defined by $t_i \in S_d(1), (d \geq 2)$, and $\theta_i \in (-1, 1)$.

1. If $n = 1$ then return `False`.

2. Let $c = (1, 0, \ldots, 0) \in \mathbb{R}^d$. Check if $c \in \partial K_i$ for some $i$. In such a case, rotate the whole hypersphere in plane $x_1 x_2$ by a small angle $\delta$ so that $c \notin \partial K_i$, $i = 1, \ldots, n$ (condition of Lemma 6).

3. Compute vectors $\beta_i$ and values $r_i$ using eq. (11). If $t_{i1} > \theta_i$ ($c \in K_i$) set $\text{out}_i = \texttt{True}$, otherwise set $\text{out}_i = \texttt{False}$. If there holds $\text{out}_i = \texttt{False}$ for every $i = 1, \ldots, n$ then return $\texttt{False}$.

4. If $d = 2$ apply Algorithm **Cover2-2D** for $\beta_{i2}$, $r_i$ and $\text{out}_i$, $i = 1, \ldots, n$ and return the obtained value. Otherwise continue.

5. For every $i = 1, \ldots, n$ do the following:

   5.1 Determine $x_j^i$ and $\theta_j^i$ (relations (18)), for every $j \neq i$, $j = 1, \ldots, n$.
   If $\theta_j^i \notin (-1, 1)$, let $A = x_j^i + (r_i, 0, \ldots, 0)$. If point $A$ belongs to $R_j$ (check the relation (14)), set $i = i + 1$ and go to step 5. Otherwise continue.
   If $\theta_j^i \in (-1, 1)$, form cap $K_j^i$, eq. (17). If $\text{out}_j = \texttt{True}$ then set $x_j^i = -x_j^i$ and $\theta_j^i = -\theta_j^i$.

   5.2 Apply Algorithm **Cover** on the set of all formed caps $K_j^i$.

Step 1 of Algorithm **Cover** implicitly covers the case when the number of caps $n$ is smaller than the number of dimensions $d$. In step 2, we introduce the rotation by a small angle $\delta$. We may set $\delta$ to an arbitrary value, for example $\delta = 0.01$. If, after the rotation, point $c$ is again on arc $C_i$, angle $\delta$ needs to be changed. One possibility is to exponentially decrease it by setting $\delta = p\delta$, where $0 < p < 1$ (we used $p = 0.9$) and to repeat the same procedure until the point $c$ is not on any arc $C_i$. Step 3 performs inversion and checks whether the caps map into external or internal regions. If all regions are internal, as discussed earlier, the coverage is $\texttt{False}$. After this, we check whether the base case of recursion is achieved. Otherwise, Step 5 checks whether the conditions of the Theorem 16 are satisfied and, when needed, performs recursive calls of **Cover**.

The worst-case time complexity of algorithm **Cover** is exponential in terms of $d$ and polynomial in terms of $n$ as shown by the following theorem:

**Theorem 8.** *The time complexity of Algorithm* **Cover** *is* $T(n, d) = \mathcal{O}\left(n^{d-1} \log n\right)$.

**Proof.**

Algorithm **Cover**, Step 5, reduces $d$-dimensional problem **SphCovVer** to at most $n$ problems of size $n - 1$ and dimension $d - 1$. The complexity of *reduce* operations (including inversion, eq. (11), and checking conditions from Step 5), is $\mathcal{O}(n^2 \cdot d)$. Hence, the following recursive relationship holds $T(n, d) = \mathcal{O}(n^2 \cdot d) + nT(n - 1, d - 1)$. Since time complexity of the base case, Algorithm **Cover2-2D**, is $\mathcal{O}(n \log n)$, this leads to the statement of the theorem. □

### 4.3 Localization of uncovered point

When the solution to **SphCovVer** problem is $\texttt{False}$, it may be of interest to identify a point on a hypersphere not covered by any of the caps. We demonstrate how this could be accomplished using results of Algorithm **Cover**.

The main idea of the proposed method is as follows. If unit hypersphere $S_d(1)$, is not completely covered by caps $K_1, \ldots, K_n$, then angles of the corresponding cones $C_i$ can be slightly widened so that the resulting system of cones still does not cover the space. Further, as demonstrated in this section, it is possible to find a point on the boundary of the region covered by the enlarged cones, which corresponds to an uncovered point of the original problem.

Namely, when the output of Algorithm **Cover** is $\texttt{False}$, we can determine a point $u \in S_d(1)$ on the boundary of the covered region. Moreover, such a point belongs to an intersection of several boundaries $D_i = \partial K_i$, but $u \notin \text{int} K_i$ for every $i = 1, \ldots, n$. If such a point belongs to exactly $l$ boundaries $D_i$, we call it $(l, d)$-*boundary* point. We propose a method for computation of the $(l, d)$-boundary point and then extend it to the computation of uncovered point. An example is illustrated in Fig. 3. The main idea is to identify a boundary point in a lower dimensional space (during recursive steps of **Cover**) and connect it with the boundary point of the original problem.

Let the output of Algorithm **Cover** be `False`. Define recursively the sequence of points $u^m \in S_m(1)$ and $v^m \in \mathbb{R}^m$ by
$$u^m = \psi_{c^m,1}(v^m), \quad v^m = \beta_i^m + r_i^m(0, u^{m-1})$$
where $m = m_0, m_0 + 1, \ldots, d$, $m_0 = \max\{d - l, 2\}$ and $i$ is the index of uncovered hypersphere in step 5 of $m$-th recursion call of Algorithm **Cover**. Also we denoted $c^m = \underbrace{(1, 0, \ldots, 0)}_{m}$ and

$$l = \begin{cases} d - 1, & \text{False is returned by Algorithm \textbf{Cover2-2D} (step 4 of \textbf{Cover})}, \\ h - 1, & \text{False is returned at steps 1 or 3 of } h\text{-th recursion call of \textbf{Cover}}. \end{cases}$$

while $\beta_i^m$ and $r_i^m$ are corresponding values $\beta_i$ and $r_i$ obtained from $m$-th recursion call of Algorithm **Cover**. In the first case $(l = 1)$, initial point $u^1$ is equal to $-1$ or $1$, depending of whether $c_i$ or $d_i$ is uncovered by sets $R_j$. Otherwise it is given by

$$u^{d-h+1} = \begin{cases} -t_1^{d-h+1}, & \text{False is returned by Step 1 of \textbf{Cover}}, \\ c^{d-h+1}, & \text{False is returned by Step 3 of \textbf{Cover}}. \end{cases} \quad (19)$$

Lemma 9, stated below, proves that $u^d$ is $(l, d)$-boundary point of the initial problem.

**Lemma 9.** *Every point $u^m$, $m = m_0, m_0 + 1, \ldots, d$, $m_0 = \min\{d - l, 3\}$ is a $(m - d + l, m)$-boundary point of the corresponding $m$-dimensional **SphCovVer** problem in $m$-th recursion call of Algorithm **Cover**.*

**Proof.** We first identify a boundary point in the base case when the `False` answer of Algorithm **Cover** is detected. Then we prove by induction that each recursive call (step 5.2 of **Cover**) results in an additional boundary to which the point belongs. There are two possibilities for the base case of induction:

1. Let the answer `False` be generated by Algorithm **Cover2-2D**. Define $u^1$ as follows
$$u^1 = \begin{cases} -1, & \text{point } c_i \text{ is uncovered} \\ 1, & \text{point } d_i \text{ is uncovered} \end{cases}$$

   It is not difficult to observe that $v^2 = (1/2, c_i)$ in the first and $v^2 = (1/2, d_i)$ in the second case. This point corresponds to point $u^2 = \psi_{c^2,1}(v^2) \in S_2(1)$ $(c^2 = (1, 0))$ which is a boundary point of some arc $K_i$ and uncovered by other arcs $K_j$, $j \neq i$. Hence, point $u^2$ is a required $(1, 2)$-boundary point for the case $d = 2$.

2. Now let the answer `False` be generated by steps 1 or 3 of Algorithm **Cover** at recursion level $h$. Point $u^{d-h+1}$ defined by eq. (19) does not belong to any hypercap (interior or boundary) and hence is uncovered. Therefore, $u^{d-h+1}$ is a $(0, d - h + 1)$-boundary point.

We prove the inductive step now. Consider the recursive call of Algorithm **Cover** on $m$-th level. If it is not otherwise stated, all notation corresponds to the **SphCovVer** problem being solved on $m$-th recursive level.fs

Assume that the recursive call in step 5.2 returned `False` on the $i$-th subproblem (i.e., covering of $S_{m-1}(1)$ by caps $K_j^i$, $j = 1, \ldots, l$). Also assume (by induction hypothesis) that returned point $u^{m-1}$ is a $(m - 1 - d + l, m - 1)$-boundary point. Point $v^m = \beta_i + r_i(0, u^{m-1})$ belongs to hypersphere $S_i$. Since $u^{m-1}$ is a $(m-1-d+l, m-1)$-boundary point, $v^m$ does not belong to the interior of any cap $K_i^k$ and hence does not belong to the interior of $R_j$ (i.e., set $R_j \setminus S_j$) for any $j$. This directly implies that $u^m = \psi_{c^m,1}(v^m)$ is not contained in the interior of any hypercap $K_j$.

Since point $u^m$ is, by assumption, contained in the boundary of $m-1-d+l$ hypercaps $K_j^i$, there exist spheres $S_{i_1}, \ldots, S_{i_{m-1-d+l}}$ so that $v^m \in S_{i_j}$ for $j = 1, \ldots, l-1$. Obviously, point $u^m$ is contained within the boundary of $l$ hypercaps $K_i, K_{i_1}, \ldots, K_{i_{m-1-d+l}}$. This completes the proof by induction that $u^m$ is a $(m-d+l, m)$-boundary point. □

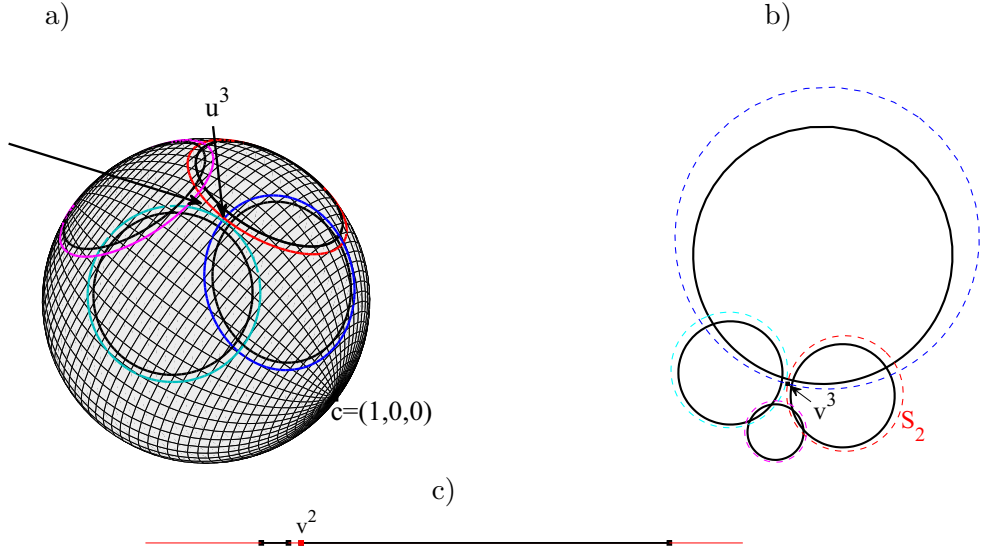

Figure 3: a) Spherical caps corresponding to cones $C_i$ with $\theta = \sqrt{3}/2$ (boundaries denoted by solid line) do not cover $S_3(1)$ and sperical caps (denoted by colored dashed lines) corresponding to cones $C_i^\alpha$ that also do not cover $S_3(1)$. b) Inversion $\psi_{(1,0,0),1}$ of spherical cap boundaries; c) Inversion $\psi_{(1,0),1}$ of 2D sphere $S_2$. Point $v^2$ is boundary point detected in Algorithm **Cover2-2D** corresponding to $(1,2)$ boundary point $u^2$ and also to point $v^3 \in S_2$ and an uncovered point $u^3$ of $S_3(1)$.

The following Lemma 10 formalizes the fact that given cones $C_i$, $i = 1, \ldots, n$ which do not completely cover space $\mathbb{R}^d$, can be enlarged (by increasing the central angle for a sufficiently small value) so that the resulting system of cones still does not cover the space.

**Lemma 10.** *If solution of **SphCovVer** problem is* `False` *for instance of cones $C_i = C(t_i; \theta_i), i = 1, \ldots, n$, then there exists sufficiently small value $\alpha$ so that' the solution of **SphCovVer** problem for cones $C_i^\alpha = C(t_i; \theta_i - \alpha), i = 1, \ldots, n$, is also* `False`.

**Proof.** Note that $\mathcal{U} = \mathbb{R}^d \setminus \bigcup_{i=1}^n C_i$ is an open set and there exists at least one internal uncovered point $x \in \mathcal{U}$. Hence there exists a ball $B_d(x; \rho)$ so that $B_d(x; \rho) \subset \mathcal{U}$. Let $\gamma_i = \arccos \theta_i$. All angles $\gamma_i$ can be enlarged by value $\Delta \gamma = 2 \arcsin(\rho/(2\|x\|))$ and corresponding enlarged cones will not contain point $x$. Hence it is sufficient to set $\alpha = \cos(\Delta \gamma)$. □

Note that $(l, d)$-boundary point $u^{d,\alpha}$ of the enlarged coverage problem (from Lemma 10) is an internal uncovered point of the original problem. This holds since the distance between point $u^{d,\alpha}$ and arbitrary cone $C_i$ is at least

$$2\|u^{d,\alpha}\| \sin\left(\frac{\arccos(\theta_i - \alpha) - \arccos(\theta_i)}{2}\right).$$

Hence, to find an uncovered point, it is sufficient to determine $\alpha$ (from Lemma 10), resolve the enlarged coverage problem and compute the boundary point (from Lemma 9). In practice, value $\alpha$ can be computed similarly as rotation angle $\delta$ (see step 2 of Algorithm **Cover**). E.g., we can set $\alpha = 0.01$ and check if cones $C_1^\alpha, \ldots, C_n^\alpha$ cover the space $\mathbb{R}^d$. If the result is True, we can exponentially decrease $\alpha$ (e.g., by setting $\alpha = p\alpha$ where $0 < p < 1$) and repeat the same procedure until the result is False. In our implementation, we choose $p = 0.9$.

Due to considerations above, the complete algorithm for computing an internal uncovered point can be formulated as follows:

**Algorithm 5. FindUncoveredPoint**
*Input: Caps $K_i$, $i = 1, \ldots, n$ defined by $t_i \in S_d(1)$ and $\theta_i \in [-1, 1]$.*

1. *Apply Algorithm* **Cover** *with values $t_i$ and $\theta_i$. If the result is* True, *return* True. *Otherwise continue.*

2. *Set $\alpha = 0.01$.*

3. *Set $\theta_i^\alpha = \theta_i - \alpha$. Apply Algorithm* **Cover** *with values $t_i$ and $\theta_i^\alpha$ and compute $(l,d)$-boundary point $u^d$.*

4. *If the result is* True, *set $\alpha = 0.9\alpha$ and go to step 2. Otherwise return* False *and point $u^d$.*

## 5 Numerical examples

In this section, we compare performance of proposed recursive algorithm **Cover** for spherical coverage verification, Section 4.2, with an algorithm based on quadratic programming **Cover-QP**, Section 4.1. We demonstrate that the application of **Cover-QP** could lead to false positives (coverage incorrectly verified) while the proposed recursive algorithm **Cover** does not suffer from such a problem. Moreover, we demonstrate that the performance of **Cover** is satisfactory in practice, in spite of its worst-case exponential complexity.

Algorithm **Cover** is implemented in programming language `C`. To implement **Cover-QP**, we utilized the programming packages `Mathematica` and `Matlab`. To test the influence of different non-convex QP solvers on the results of the algorithm, we also created an AMPL model [9] for the Algortihm **Cover-QP** and tested it using `MINOS` and `FortMp` solvers.

Implementations are tested on several test examples. In the experiments, hypercaps are determined by constellations $t_i, i = 1 \ldots, n$ of points and cones have a constant angle, i.e., $\theta_i = \theta$. This stipulation comes from applications in methods for finding inverse $k$-nearest neighbors. Namely, an algorithm for reverse $k$-nearest neighbor problem from [1, 15] requires the covering constellation with minimal $n$ and $\theta = \cos(\pi/6) = \sqrt{3}/2$.

Test constellations are generated by the relaxation algorithm from [14]. This algorithm produces near-uniform placement of the points on $d$-dimensional hypersphere. It starts with a randomly generated set of initial points, where points interact through generalized electromagnetic interactions and each point has equal charge. The algorithm seeks the solution of the $d$-dimensional generalization of the Thompson's problem [2], and searches iteratively for the equilibrium state (the state with minimal electrostatic energy).

### 5.1 Accuracy of algorithms

Our experimental results indicate that Algorithm **Cover-QP**, which utilizes solvers for concave QP problems, can be numerically very unstable. As a consequence, the result is a potentially large number of false positives (an algorithm falsely indicates that a sphere is covered by caps).

Consider constellation `four_D_85` obtained by relaxation algorithm for $d = 4$ and $n = 85$ (the whole constellation is given in the Appendix and can be found at `tesla.cis.desu.edu/data/Constellations`).

`Matlab` implementation of the Algorithm **Cover-QP** returns `True`. We tested `Mathematica` implementations for different working precisions (double precision, 20 digits, 50 digits and 100 digits). In all testings, result was `True`, but the corresponding optimal point and objective function values were different (see the following table). Results of testing AMPL model were similar.

|  | Objective function value | Optimal point | | | |
| :---: | :---: | :---: | :---: | :---: | :---: |
|  |  | $x_1$ | $x_2$ | $x_3$ | $x_4$ |
| `Matlab`: double precision | 0.9756 | 0.9477 | -0.0063 | -0.0004 | 0.2314 |
| `Mathematica`: double precision | 0.935851 | 0.172452 | 0.898297 | -0.158036 | -0.272394 |
| `Mathematica`: 20 digits | 0.891469 | -0.390757 | 0.136777 | 0.77625 | 0.342789 |
| `Mathematica`: 50 digits | 0.946319 | 0.652345 | -0.476475 | 0.236943 | 0.487436 |
| `Mathematica`: 100 digits | 0.958959 | -0.901428 | 0.034657 | -0.231813 | 0.302405 |
| `MINOS` | 0.953615 | 0.628285 | 0.469643 | 0.426044 | -0.39597 |
| `FortMP` | 0.947382 | 0.037172 | 0.869515 | 0.0528789 | 0.224396 |

This example clearly shows the instability of Algorithm **Cover-QP**. One reason for such instability may be the fact that optimization heuristics embedded in the QP solver traps into the local maximum and do not achieve the global maximum. Such a conclusion is supported by the fact that the optimal solution found by the solver changes when simple variable transformation $x = x' + c$ ($c \in \mathbb{R}^d$ is a constant vector) is applied. For example, if we put $c = (0.1, 0.4, -0.7, -0.3)$ (this choice comes from the uncovered point shown in the next paragraph) all solvers return objective function value $1.000409053$ and the optimal point $(0.134335, 0.4576476, -0.7915343, -0.3826162)$. Since the objective function value is greater than 1, in such a case algorithm returns `False`.

Note that Algorithm **Cover-QP** can have only false positives (for a non-covering constellation, providing answer `True`), not false negatives. This can be explained by the fact that a point returned from the QP solver is always a feasible solution and the objective function value in this point cannot be larger than the global maximum.

Unlike **Cover-QP**, Algorithm **Cover** is not based on heuristics and randomized initial conditions, but on relatively simple and numerically stable algebraic operations. During our experiments, we could not identify any case where **Cover** would return incorrect results. Particularly, on the same test example as above (the constellation `four_D_85`) **Cover** correctly returns `False`, and detects the following uncovered point:

$$x^{\text{uncovered}} = (0.134309, 0.457496, -0.791181, -0.383002).$$

Since $\max_{1 \leq i \leq 85}(x^{\text{uncovered}}, t_i) = 0.865901 < \cos(\pi/6)$ (see the Appendix for vectors $t_i$ of the constellation `four_d_85`) we can verify that $x^{\text{uncovered}}$ is indeed uncovered.

Note that the result of the Monte-Carlo based approach method (see Section 4) on `four_D_85` with $N = 10^{10}$ pseudo-random points was also incorrect (i.e., `True`). This again demonstrates why Algorithm **Cover** seems to be the only reliable method for spherical coverage verification.

## 5.2 Performance of recursive algorithm

We test the working time of the Algorithm **Cover**. Implementation is compiled by a GNU C compiler and runs on the machine with AMD Phenom II CPU on 3.0 GHz and CentOS 5.2 (Linux) operating system. Testing is performed for $d = 3, 4, 5$ and $200 \leq n \leq 500$. Results are shown in the Fig. 4. Execution times shown are obtained by averaging through 20 runs of the program on different constellations of the same dimensionality. All testing constellations corresponded to a sphere covered by caps (output of the Algorithm **Cover** was `True`). This restriction is added since the working time on constellations without coverage is considerably smaller due to the fact that Algorithm **Cover** does not complete recursion.

As it can be seen from the graph, Algorithm **Cover** is practically applicable and does not reach its theoretically obtained complexity. This can be explained by the fact that not all $d - 1$-dimensional hyperspheres $S_i$ intersect (Case 1 of subsection 4.1 holds). Practically, the number of hyperspheres intersecting $S_i$ is drastically smaller than $n$, which implies that the corresponding subproblems have a lower dimensionality.

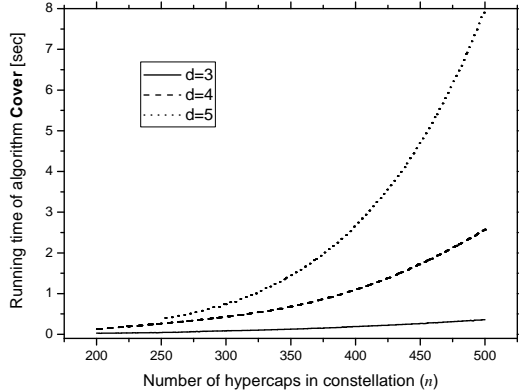

Figure 4: Average running time of Algorithm **Cover** for $d = 3, 4, 5$.

The applicability of Algorithm **Cover** made possible to determine the upper bound $\mathbf{M}_u(d)$ of minimal number $\mathbf{M}(d)$ of hypercaps with $\theta = \cos(\pi/6)$, covering the unit hypersphere. In [15], it is proved that $\mathbf{M}(d)$ is at least $\mathbf{M}_l(d) = \Theta(2^d \sqrt{d})$, i.e., it is exponential as the function of $d$. In order to obtain $\mathbf{M}_u(d)$, we generate several constellations for fixed value $n$. If the covering constellation is found, $n$ is decreased by 1 and the whole procedure is repeated. Otherwise we put $\mathbf{M}_u(d) := n$. The following table shows the values of $\mathbf{M}_u(d)$ for $d = 3, 4, 5, 6$:

| $d$ | 3 | 4 | 5 | 6 |
|---|---|---|---|---|
| $\mathbf{M}_u(d)$ | 22 | 81 | 234 | 715 |

However, the closed-form expression for upper bound $\mathbf{M}_u(d)$ is still unknown and its existence is left as an open question.

# 6 Conclusion

We have considered the spherical coverage problem: given a set of hypercaps in $d$-dimensional space, determine whether a $d$-dimensional hypersphere is completely covered by the hypercaps. We have demonstrated that the considered problem is NP hard by reducing concave quadratic programming (QP) problem to it. We have discussed two algorithms to resolve the spherical coverage problem: the first method (Algorithm **Cover-QP**) is based on the utilization of quadratic programming. The second method (Algorithm **Cover**) is recursive and based on the reduction of the main problem on $\mathcal{O}(n)$ problems of dimension $d - 1$. The recursive algorithm also provides a method to determine an uncovered point (if such a point exists).

While the worst-case time complexity of the proposed recursive method is $\mathcal{O}\left(n^{d-1} \log n\right)$, it is of practical interest, due to NP hardness of the considered problem (note also that the asymptotic complexity could be improved if, as the base case, we choose the case $d = 3$ of problem **HpCovVer** and utilize a method proposed in [4] to resolve it). However, numerical experiments indicate that the recursive algorithm almost never reaches maximal complexity and hence typically does not require prohibitive execution time. In contrast, Algorithm **Cover-QP**, with heuristics-based concave QP problem solvers, can be numerically unstable resulting in false positive detection (indicating false coverage of the sphere) and hence having limited practical application.

Our results indicate that the recursive algorithm may be the best algorithm for the problems having a relatively small dimension $d$. For the high values of $d$, where direct application of the recursive algorithm may be too time consuming, quadratic programming method still could be used as a presolve method, since it does not have false negatives.

We conclude the paper with the following open problem:

**Problem 5.** *Develop an efficient algorithm for construction of the covering constellation of minimal size, for a given dimension $d$ and angle $\theta$.*

# Acknowledgement

The study is supported by the US Department of Transportation Office of the Secretary Grant No. DTOS59-08-G-0014. M. Petković is supported by the research project 174013 of the Serbian Ministry of Science and DoD/DoA (award 45395-MA-ISP). D. Pokrajac has also been partially supported by NIH (grant #2 P20 RR016472-04), DoD/DoA (awards 45395-MA-ISP, P-54412-CI-ISP) and NSF (awards #0320991, CREST grant (#HRD-0630388 CREOSA).). Part of this research was done while M. Petković was visiting Delaware State University, Department of Computer and Information Sciences.

Authors extend their special thanks to Dr. Stephen A. Vavasis, professor and university research chair of University of Waterloo, Ontario, Canada, for help in formulating proof for NP completeness of the concave QP problem discussed in the paper. Dr. Richard McCallister, an associate professor of Delaware State University English and Foreign Languages Department, helped us in proofreading the manuscript and improving its English. In addition, authors wish to thank to Samantha McDaniel for help on drawing the figures. Also, authors are thankful to anonymous reviewers whose inputs significantly helped improving the quality of the manuscript.

# References

[1] J. Anderson, B. Tjanden, *The inverse nearest neighbor problem with astrophysical applications*, Proceedings of the 12th Symposium of Discrete Algorithms (SODA), Washington, DC, January 2001.

[2] N. Ashby, W.E. Brittin, *Thomson's Problem*, Amer. J. Phys. 54, pp 776–777, 1986.

[3] G. Barequet, M. T. Dickerson, Y. Scharf, *Covering points with a polygon*, Comput. Geom. 39 (2008), 143-162.

[4] F. Cazals, S. Loriot, *Computing the Arrangements of Circles on a Sphere with Applications in Structural Biology*, Comput. Geom. 42:6-7 (2009), 551–565.

[5] S. Cabello, J. M. Diaz-Banez, C. Seara, J. A. Sellares, J. Urrutia, I. Ventura, *Covering point sets with two disjoint disks or squares*, Comput. Geom. 40 (2008), 195-206.

[6] T. H. Cormen, C. E. Leiserson, R. L. Rivest, C. Stein *Introduction to Algorithms, 2nd edition*, The MIT Press, Boston, 2001.

[7] I. Dumer, *Covering Spheres with Spheres*, Discrete and Computational Geometry 38, 665-679 (2007).

[8] K. Elbassioni, H.R. Tiwary, *On a cone covering problem*, Comput. Geom. (2010), doi: 10.1016/j.comgeo.2010.07.004.

[9] R. Fourer, D.M. Gay, B.W. Kernighan, *AMPL: A Modeling Language for Mathematical Programming*, Duxbury Press, Brooks/Cole Publishing Company, 2002.

[10] R.M. Freund, J.B. Orlin, *On the complexity of four polyhedral set containment problems*, Mathematical programming 33 (1985), 139–145.

[11] J. Han, M. Kamber, *Data Mining: Concepts and Techniques*, Morgan Kaufmann Publishers, 2001.

[12] F.J. MacWilliams, N.J.A. Sloane, *The theory of error-correcting codes*, North-Holland mathematical library, 2006.

[13] D. Pokrajac, A. Lazarevic, L.J. Latecki, *Incremental Local Outlier Detection for Data Streams*, Proceedings IEEE Symposium on Computational Intelligence and Data Mining, CIDM 2007, pp 504–515.

[14] D. Pokrajac, J. Milutinović, I. Ekanem, *Application of electrostatic method for uniform placement of points on a hypersphere*, Proceedings of the international conference of applied electromagnetics - PES 2007, Niš, Serbia.

[15] D. Pokrajac, M.D. Petković, L.J. Latecki, A. Lazarević, N. Reljin, J. Milutinović, *Computational Geometry Issues of Reverse Nearest Neighbor Algorithm*, Hawaii International Conference on Statistics, Mathematics and Related Fields, Honolulu, HI, January 2008.

[16] D. Pokrajac, N. Reljin, N. Pejčić, A. Lazarević, *Incremental Connectivity-Based Outlier Factor Algorithm*, Proc. Visions in Computer Science, 2008.

[17] S.A. Vavasis, *Nonlinear Optimization: Complexity Issues*, Oxford University Press, USA, 1991.

[18] J.-L. Verger-Gaugry, *Covering a Ball with Smaller Equal Balls in $\mathbb{R}^n$*, Discrete and Computational Geometry 33, 143-155 (2005).

# A  Constellation `four_D_85`

| $t_1$ | $t_2$ | $t_3$ | $t_4$ |
|---|---|---|---|
| 0.911722 | 0.083517 | -0.402106 | 0.009974 |
| 0.581301 | -0.242364 | 0.278539 | -0.725096 |
| 0.218664 | 0.316284 | 0.287611 | 0.877172 |
| 0.117213 | 0.928431 | -0.32201 | 0.143484 |
| -0.737931 | -0.158832 | -0.035717 | 0.654946 |
| 0.36765 | -0.515669 | -0.616088 | 0.468352 |
| 0.76061 | 0.150628 | -0.006887 | 0.631456 |
| -0.321449 | -0.24082 | 0.870032 | 0.285869 |
| 0.344644 | 0.293037 | 0.546223 | -0.704976 |
| -0.412761 | 0.540358 | 0.013305 | -0.73312 |
| 0.816541 | -0.164327 | 0.467615 | 0.295963 |
| -0.245525 | -0.585043 | -0.116853 | 0.76406 |
| 0.338649 | -0.178681 | -0.219111 | 0.89743 |
| -0.69018 | -0.518921 | 0.382654 | 0.328555 |
| 0.740476 | 0.306078 | 0.049102 | -0.596322 |
| -0.706023 | -0.187694 | 0.659171 | -0.178316 |
| -0.055919 | -0.589895 | -0.805087 | -0.027052 |
| 0.328685 | -0.068095 | 0.878784 | 0.339218 |
| 0.943282 | -0.273977 | 0.075663 | -0.171553 |
| -0.159834 | 0.605736 | 0.643967 | -0.43914 |
| 0.059374 | 0.684132 | 0.716063 | 0.125265 |
| 0.562506 | 0.316235 | 0.756618 | -0.105413 |
| 0.526948 | -0.692313 | -0.153046 | -0.468621 |
| 0.585505 | -0.29157 | 0.718579 | -0.236252 |
| -0.339912 | -0.865955 | -0.293451 | 0.220156 |

| $t_1$ | $t_2$ | $t_3$ | $t_4$ |
|---|---|---|---|
| 0.504902 | 0.689107 | -0.350765 | -0.383627 |
| -0.418404 | 0.437219 | 0.203124 | 0.769752 |
| 0.498145 | -0.17126 | -0.337765 | -0.780023 |
| -0.19152 | -0.885608 | -0.106055 | -0.409599 |
| -0.216553 | 0.126561 | -0.29997 | 0.920383 |
| 0.060341 | 0.820685 | 0.197331 | 0.532819 |
| -0.736942 | -0.41477 | -0.460301 | 0.270194 |
| -0.547833 | -0.450474 | -0.569484 | -0.415501 |
| 0.204308 | -0.777431 | 0.433251 | -0.407619 |
| -0.983534 | -0.127577 | 0.121635 | 0.039876 |
| -0.106398 | 0.780104 | -0.465982 | -0.403706 |
| -0.824089 | 0.406231 | 0.311938 | -0.241967 |
| 0.925197 | 0.302318 | 0.226185 | -0.038143 |
| 0.582566 | 0.128159 | -0.589217 | 0.544991 |
| 0.622845 | -0.704481 | 0.326492 | 0.095782 |
| -0.836308 | 0.452032 | -0.011796 | 0.310027 |
| -0.309656 | 0.548631 | -0.775807 | 0.035212 |
| -0.225897 | -0.143505 | 0.308569 | 0.912777 |
| 0.656179 | -0.438453 | -0.607877 | -0.087605 |
| -0.12433 | -0.638736 | 0.476572 | 0.591132 |
| -0.425256 | 0.298745 | 0.854262 | -0.012046 |
| 0.480447 | 0.819603 | 0.259428 | -0.173544 |
| -0.514513 | 0.096686 | 0.503361 | -0.687428 |
| 0.267199 | 0.5784 | -0.305584 | 0.707585 |
| -0.018035 | -0.020976 | -0.880919 | -0.472457 |
| -0.736443 | -0.653951 | -0.063084 | -0.161308 |
| 0.033759 | -0.030823 | 0.959959 | -0.276386 |
| -0.471801 | 0.343863 | -0.601331 | -0.545493 |
| -0.349645 | 0.727398 | -0.321563 | 0.495215 |
| 0.004942 | -0.96662 | 0.232848 | 0.106783 |
| 0.111029 | -0.625644 | 0.768139 | 0.078766 |
| 0.060656 | -0.490269 | -0.013342 | -0.869356 |
| 0.588121 | 0.537068 | 0.416267 | 0.438626 |
| 0.338081 | -0.727789 | 0.016177 | 0.596458 |
| -0.342943 | -0.152519 | -0.336818 | -0.863529 |
| -0.868041 | 0.131346 | -0.46866 | -0.098029 |
| 0.084497 | -0.533808 | -0.58593 | -0.603817 |
| 0.555933 | 0.174623 | -0.695937 | -0.419663 |
| 0.409128 | -0.265222 | 0.414691 | 0.768312 |
| -0.398058 | -0.719619 | 0.526168 | -0.216435 |
| -0.230651 | -0.326573 | -0.707467 | 0.582787 |
| -0.629986 | 0.173969 | -0.5524 | 0.517404 |
| -0.389677 | -0.043493 | -0.919642 | 0.0228 |
| -0.013034 | 0.298631 | -0.792652 | 0.531369 |
| 0.186482 | 0.689421 | 0.091131 | -0.693987 |
| 0.814852 | -0.386301 | -0.176673 | 0.39443 |
| 0.275837 | -0.088184 | -0.946315 | 0.143616 |
| -0.497203 | -0.491744 | 0.211614 | -0.682786 |
| -0.11172 | 0.959906 | 0.17654 | -0.186902 |
| -0.028234 | 0.117661 | 0.115501 | -0.98591 |
| 0.083739 | 0.314023 | -0.444543 | -0.834721 |
| -0.805373 | -0.010126 | -0.060896 | -0.589545 |
| -0.125314 | 0.270743 | 0.737091 | 0.606376 |
| 0.690523 | 0.637669 | -0.183595 | 0.287835 |
| -0.481723 | 0.736347 | 0.436683 | 0.1872 |
| -0.719834 | 0.10622 | 0.559014 | 0.397567 |
| 0.38002 | 0.543697 | -0.745668 | 0.062905 |
| -0.653788 | 0.707026 | -0.192191 | -0.189048 |

| $t_1$ | $t_2$ | $t_3$ | $t_4$ |
|---|---|---|---|
| 0.013797 | -0.290263 | 0.625477 | -0.72411 |
| 0.303814 | -0.890275 | -0.328701 | 0.084045 |

## B  Pseudocode of Algorithm Cover

Here we give the complete pseudo-code description of Algorithm **Cover**.

**Algorithm 1** $\mathbf{Cover}(n; d; \mathbf{X}; \theta)$ - Spherical coverage verification

**Require:** Centers and angles of cones: $\mathbf{X} = [t_1^T \cdots t_n^T]^T \in \mathbb{R}^{n \times d}$ and $\theta = (\theta_1, \ldots, \theta_n)$
1: **if** $d = 2$ **then**
2:     **return** $\mathbf{Cover2}(n; \mathbf{X}; \theta)$
3: **end if**
4: $S := (1, 0, \ldots, 0) \in \mathbb{R}^d$
5: **if** $(S, t_i) = \theta_i$ for some $i = 1, 2, \ldots, n$ **then**
6:     $\mathbf{X} := \mathbf{Rot}(n; d; \mathbf{X}; \theta)$ (rotates by small angle $\delta$).
7: **end if**
8: $r_i := \frac{\sqrt{1-\theta_i^2}}{2(\theta_i - t_{i1})}$ for $i = 1, \ldots, n$ (radii of the spheres in $d-1$-dimensional hyperplane)
9: $\beta_{ij} := \frac{t_{ij}}{2(\theta_i - t_{i1})}$ for $i = 1, \ldots, n$ and $j = 2, \ldots, d$ (centers of the spheres in $d-1$-dimensional hyperplane)
10: $\text{out}_i := ((S, t_i) < \theta_i)$ (True if $S$ belongs to the interior of cap $K_i$ and False otherwise)
11: **for** $i := 1$ to $n$ **do**
12:     $k := 0$
13:     $\text{Cover}_i := \text{False}$
14:     **for** $j := 1$ to $n$ **do**
15:         **if** $i \neq j$ **then**
16:             $k := k + 1$
17:             $t_{k,l-1}^1 := \beta_{jl} - \beta_{il}$ for $l = 2, 3, \ldots, d$
18:             $d_{ij} := \|t_k^1\|$
19:             $\theta_k^1 := \frac{r_i^2 + d_{ij}^2 - r_j^2}{2 r_i d_{ij}}$
20:             **if** $-1 \leq \theta_k^1 \leq 1$ **then**
21:                 **if** $\text{out}_j$ **then**
22:                     $\theta_k^1 := -\theta_k^1$
23:                     $t_{kl}^1 := -t_{kl}^1$ for $l = 1, 2, \ldots, d-1$
24:                 **end if**
25:             **else**
26:                 $s := \sqrt{(\beta_{i2} + r_i - \beta_{j2})^2 + \sum_{l=3}^{d}(\beta_{il} - \beta_{jl})^2}$
27:                 **if** ($s \leq r_j$ and not $\text{out}_j$) or ($s \geq r_j$ and $\text{out}_j$) **then**
28:                     $\text{Cover}_i := \text{True}$ ($j$-th hypersphere completely covers $i$-th)
29:                     break
30:                 **else**
31:                     $k := k - 1$ ($j$-th hypersphere is disjoint from $i$-th)
32:                 **end if**
33:             **end if**
34:         **end if**
35:     **end for**
36:     **if** not $\text{Cover}_i$ **then**
37:         **if** $k = 0$ **then**
38:             **return** False
39:         **else**
40:             $\mathbf{X}^1 := [(t_1^1)^T \cdots (t_n^1)^T]^T \in \mathbb{R}^{k \times d-1}$
41:             $\theta^1 := (\theta_1^1, \ldots, \theta_k^1)$
42:             $\text{Cover}_i := \mathbf{Cover}(k; d-1; \mathbf{X}^1; \theta^1)$
43:             **if** not $\text{Cover}_i$ **then**
44:                 **return** False
45:             **end if**
46:         **end if**
47:     **end if**
48: **end for**
49: **return** True

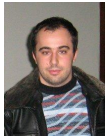
**Marko D. Petković** was born in Niš, Serbia, in 1984. He graduated in mathematics and computer science at the Faculty of Science and Mathematics, Niš, Serbia in 2006. He also graduated in telecommunications at Faculty of Electronic Engineering, Nis, Serbia in 2007. Also he received his PhD degree in Computer Science from the University of Niš in 2008. Currently he is an Assistant Professor in Computer Science at the Faculty of Science and Mathematics, University of Niš, Serbia. His research interests include the theory and computation of matrix generalized inverses, computational geometry, source and channel coding, symbolic computation and optimization methods. He is the author of about 50 papers (about 30 of them in peer-reviewed international journals).

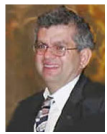
**Dragoljub Pokrajac** was born in Šibenik, Croatia. He received his BS in electrical engineering (1993) and MS in telecommunications at University of Niš, Serbia (1997). He attended the PhD program at Washington State University 1998-2000 and received his PhD in Computer Science from Temple University in 2002. He is currently a tenured Associate Professor at Delaware State University, Dover, DE. Also, he worked as a senior scientist at VIP Mobile, Inc. 2008-2010. His research interests include machine learning, computer vision, video analytics, computational geometry and theory of algorithms. Dr. Pokrajac is member of ACM and F&AM Equity Lodge 591, Pennsylvania.

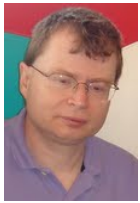
**Longin Jan Latecki** received the PhD degree in computer science from Hamburg University, Germany, in 1992. He is a Professor of Computer Science at Temple University, Philadelphia. His main research interests include shape representation and similarity, object detection and recognition in images, robot perception, data mining, and digital geometry. He has published over 175 research papers and books. He is an editorial board member of Pattern Recognition and International Journal of Mathematical Imaging. He received the annual Pattern Recognition Society Award together with Azriel Rosenfeld for the best article published in the journal Pattern Recognition in 1998.